# Block Bayesian Sparse Learning Algorithms With Application to Estimating Channels in OFDM Systems

Guan Gui and Li Xu
Department of Electronics and Information
Systems, Akita Prefectural University,
Akita, 015-0055, Japan
Emails: {guiguan, xuli}@akita-pu.ac.jp

Lin Shan
Wireless Network Research Institute, National Institute
of Information and Communications Technology
Yokosuka, 239-0847, Japan
Email: shanlin@nict.go.jp

*Abstract*—Cluster-sparse channels often exist in frequency-selective fading broadband communication systems. The main reason is received scattered waveform exhibits cluster structure which is caused by a few reflectors near the receiver. Conventional sparse channel estimation methods have been proposed for general sparse channel model which without considering the potential cluster-sparse structure information. In this paper, we investigate the cluster-sparse channel estimation (CS-CE) problems in the state of the art orthogonal frequency-division multiplexing (OFDM) systems. Novel Bayesian cluster-sparse channel estimation (BCS-CE) methods are proposed to exploit the cluster-sparse structure by using block sparse Bayesian learning (BSBL) algorithm. The proposed methods take advantage of the cluster correlation in training matrix so that they can improve estimation performance. In addition, different from our previous method using uniform block partition information, the proposed methods can work well when the prior block partition information of channels is unknown. Computer simulations show that the proposed method has a superior performance when compared with the previous methods.

## I. INTRODUCTION

Frequency-selective fading channels are caused by the reflection, diffraction and scattering of the transmitted signals due to the buildings, moving vehicles, mountains, etc [1]. Such fading phenomenon distorts received signals and poses critical challenges in the design of communication systems for high-rate and high-mobility wireless communication applications. Hence, accurate channel state information is required at receiver for coherent detection [2].

In recent years, many channel measurement experiments verified that wireless channel exhibits sparse or/and cluster-sparse structure (as shown in Fig. 1) since broadband signal transmission. For example, many real-world channels of practical interest, such as in underwater acoustic communication [3], terrestrial signals transmission of high definition television (HDTV) [4] and residential ultra wideband (UWB) systems [5]–[7], tend to have sparse or approximately sparse impulse responses. For sparse channel model based communication systems, various spectral-efficient sparse channel estimation methods have been proposed [8]–[13]. It was well known that in many propagation environments, there exist several big obstacles, e.g., buildings and hilly-terrains environment, which give rise to cluster-sparse structure in multipath channels [14], [15]. One of typical examples is shown in Fig. 1, where the channel delay-spread sampled length is 80 and active clusters are 2. Note that each active cluster is consisted of 10 nonzero taps. Given the block partition, a number of algorithms, such as group Lasso [16], block orthogonal matching pursuit (OMP) [17], and block compressive sampling matching pursuit (CoSaMP) [18] have been proposed for taking advantages of the block-structure information in signals. For example, BOMP has been successfully applied in the cluster-sparse channel information [10].

However, the proposed method assumed that channels can be split uniformly and employed this prior information on CS-CE. In real communication systems, it is impossible to get the partition prior information on channel estimation. So, how to adaptively estimate the cluster partition while recovering the sparse channel is a challenge issue.

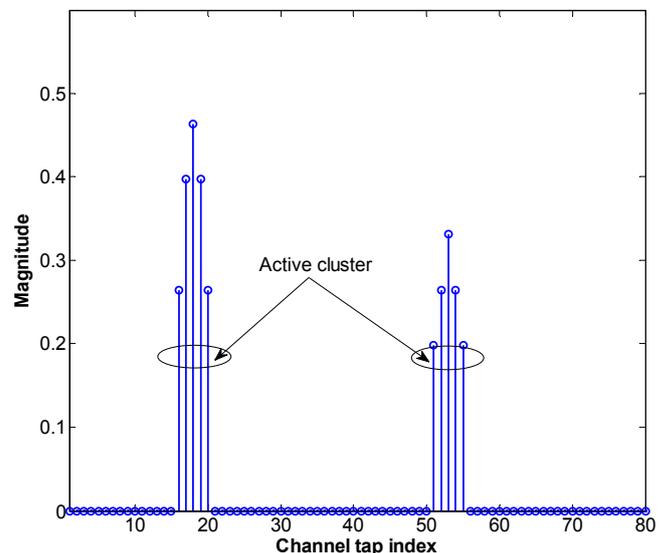

Fig. 1. A typical example of cluster-sparse channel model.

We first assume that all the nonzero blocks are of equal size $h$ and are arbitrarily located. This model is consistent with communication channel modeling where an ideal sparse channel consisting of a few specular multi-path components has a discrete-time, bandlimited, baseband representation, which exhibits a cluster sparse structure with the cluster centers determined by the arbitrary arrival times of the multi-path components. Since the clusters are arbitrarily located they can overlap giving rise to larger unequal blocks, making the model quite flexible. So, the assumption of equal block-size is not limiting.

One can find that traditional methods applied either sparse channel model or cluster-partition information. Motivated by the recent block sparse Bayesian learning (BSBL) algorithms [19], which can learn and exploit intra-block correlation and have superior performance to above state-of-the-art algorithms. Unlike traditional CS-CE methods, in this paper, we propose two Bayesian cluster-sparse channel estimation (BCS-CE) methods by adopting BSBL algorithms, to exploit both cluster-structure and sparse-structure information in OFDM channels. The performance of the proposed methods is validated by computer simulation via mean square error (MSE) performance and computational complexity (CPU time).

This paper is organized as follows. Section II introduces OFDM channel model and problem formulation. To estimate the cluster-sparse channel, in section III, we propose BCS-CE methods by using BSBL algorithm with block partition information. In section IV, we propose second BCS-CE method without utilizing the partition information. In section V, several typical computer simulation results are given to verify the effectiveness of our proposed method. Section VI concludes with a discussion on related work and future research.

## II. SYSTEM MODEL AND PROBLEM FORMULATION

Let the OFDM symbol is consisted of $N_d$ subcarriers and pilots are allocated $N$ subcarriers for acquiring channel state information. In addition, assume that the length of the cyclic prefix (CP) in the OFDM symbols is greater than $\tau_{max} \geq \tau_l$, $l = 0, 1, \cdots, N-1$. Suppose that $X(i)$ denote the $i$-th subcarrier in an OFDM symbol, where $i = 0, 1, \cdots, N_d - 1$. If the coherence time of the channel is much larger than the OFDM symbol duration $T$, then the channel can be considered static over an OFDM symbol. Let $\bar{y}$ be the vector of received signal samples in one OFDM symbol after DFT, then

$$\bar{y} = \bar{X}\bar{h} + \bar{z} = \bar{X}Fh + \bar{z}, \quad (1)$$

where $\bar{X} = \text{diag}\{X(0), X(1), \cdots, X(N-1)\}$ denotes diagonal subcarrier matrix; $\bar{h}$ is the channel frequency response (CFR) in frequency-domain (FD); $\bar{z}$ is noise vector; and $F$ is a $N \times L$ partial DFT matrix with its $k$-th row vector $1/\sqrt{N}[1, e^{-j2\pi k/N}, \cdots, e^{-j2\pi k(N-1)/N}]^T$. Since $\bar{h} = Fh$, the FD channel response $\bar{h}$ lies in the time-delay spread domain, the $L$-length discrete channel vector can be written as $h = [h_0, h_1, \cdots, h_{L-1}]^T$. By means of virtual block partition, the channel vector $h$ can also be written clustered-sparse form as

$$h = [\underbrace{h_1, \cdots, h_d}_{h_1^T}, \cdots, \underbrace{h_{d(i-1)+1}, \cdots, h_{di}}_{h_i^T}, \cdots, \underbrace{h_{dc-1}, \cdots, h_{dc}}_{h_c^T}]^T, \quad (2)$$

where $L = cd$, $c$ is the total partition number of channel clusters and $d$ is the length of each cluster. Our objective is to estimate cluster-sparse channel $h$ by using a observed signal $\bar{y}$ and an equivalent training signal matrix $X = \bar{X}F$.

## III. BCS-CE WITH THE CLUSTER-PARTITION INFORMATION

Considering the cluster-sparse channel model in (2), we assume that each channel cluster $h_i$ satisfies a parameterized multivariate Gaussian distribution as

$$P(h_i) \sim \mathcal{CN}(0, \gamma_i B_i), \quad i = 1, 2, \cdots, c, \quad (3)$$

where $\gamma_i \geq 0$ is a parameter which controls the number of active channel clusters, i.e., $\gamma_i = 0$ and $\gamma_i > 0$ denote $i$-th block as zero and active cluster, respectively. During the Bayesian channel estimation procedure, fortunately, most of parameters $\gamma_i$, $i = 1, 2, \cdots, c$, tend to zero or inactive in real wireless channels due to multipath correlations. Please notice that $B_c$ is adopted to capture correlation channel structure information of the $i$-th cluster. In addition, we assume that the clusters are mutually uncorrelated in $h$, given by $P(h) \sim \mathcal{CN}(0, D)$, where $D$ is a block-diagonal matrix with each principal block given by $\gamma_i B_i$. The noise vector $\bar{z}$ is assumed to satisfy mean and variance as $P(\bar{z}) \sim \mathcal{CN}(0, \lambda I)$, where $\lambda \geq 0$ is to control noise level. Hence, the posterior information of cluster channel vector $h$ can be computed by

$$P(h|y; \lambda, \{\gamma_i, B_i\}_{i=1}^c) \sim \mathcal{CN}(\mu_h, D_h), \quad (4)$$

where $\mu_h$ and $D_h$ are given by

$$\mu_h = DX^T(\lambda I + XDX^T)^{-1}\bar{y}, \quad (5)$$

$$D_h = (D^{-1} + 1/\lambda X^T X)^{-1}, \quad (6)$$

respectively. According to above Eqs. (4)~(6), once the hyper-parameters $\lambda$, $\{\gamma_i, B_i\}_{i=1}^c$ are estimated, the maximum a posterior estimate of $h$ can be directly obtained from the mean of the posterior [20]. By adopting the BSBL [19] and utilizing the cluster partition information, the first BCS-CE method is summarized in Tab. 1.

Above proposed BCS-CE method in Tab. 1, cluster-partition information was assumed to know. In other words, same $B$ was utilized for all of channel partition blocks. If we constraint the $B$, the estimation performance can be further improved. This is motivated by the fact that we can find a positive definite and symmetric matrix $\tilde{B}$ such that $\tilde{B}$ is close to $B$ especially in the elements along the main diagonal and the main sub-diagonal. A possible form of $\tilde{B}$ is given by

$$\tilde{B} \sim \text{Toeplitz}\left([1, r, \cdots, r^{d-1}]\right)$$
$$= \begin{bmatrix} 1 & r & r^2 & \cdots & r^{d-1} \\ r & 1 & r & \cdots & r^{d-2} \\ \vdots & \vdots & \vdots & \ddots & \vdots \\ r^{d-1} & r^{d-2} & r^{d-3} & \cdots & 1 \end{bmatrix}, \quad (7)$$

with $r \triangleq \text{sign}(m_1/m_0) \min\{|m_1/m_0|, 0.9\}$, where $m_0$ is the average of the elements along the main diagonal and $m_1$ is the average of the elements along the main sub-diagonal of the matrix $B$ in Tab. 1. 0.9 is a bound to control the $r$.

TABLE I. PROPOSED BCS-CE METHOD USING CLUSTER-PARTITION INFORMATION.

| |
|---|
| Input: $\bar{X}$, $\bar{y}$, $F$, $D$ |
| Output: $\tilde{h}$ |
| Initialize: $\mu_h \leftarrow 0$, $D_h \leftarrow 0$, <br> For $i = 1, 2, \ldots, c$ <br> $\quad B \leftarrow \frac{1}{c} \sum_{i=1}^{c} \frac{D_h^i + \mu_h^i (\mu_h^i)^T}{\gamma_i}$ <br> $\quad B_i \leftarrow B$ <br> $\quad \gamma_i \leftarrow \frac{1}{d_i} Tr\left[ B^{-1}(D_h^i + \mu_h^i(\mu_h^i)^T) \right]$ <br> $\quad \mu_h \leftarrow DX^T(\lambda I + XDX^T)^{-1}y$ <br> $\quad D_h \leftarrow (D^{-1} + 1/\lambda X^T X)^{-1}$ <br> $\quad \lambda \leftarrow \frac{\|\bar{y} - X\mu_h\|_2^2 + \lambda\left[L - \text{Tr}(D_h D^{-1})\right]}{N}$ <br> End <br> $\tilde{h} \leftarrow DX^T(\lambda I + XDX^T)^{-1}\bar{y}$ |

## IV. BCS-CE WITHOUT CLUSTER-PARTITION INFORMATION

In many scenarios, the cluster partition information in channel model (2) is not available. Here we generalize the conventional cluster sparse channel model with the unknown cluster partition information. Based on the model, we adopt another BSBL algorithm [19] on BCS-CE based channel estimation, which requires little a priori knowledge on the cluster structure. Please note that based on this generalized model, many existing algorithms can also be used. Given the identical cluster size $d$, there are $k = L - d + 1$ possible clusters in $h$, which overlap each other. The $i$-th block starts at the $i$-th element and ends at the $(i + d - 1)$-th element of $h$. All the nonzero taps in $h$ are assumed to lie in some of these clusters. Similar to Section III, for the $i$-th cluster, we assume it satisfies a multivariate Gaussian distribution with mean given by $0$ and covariance matrix given by $\gamma_i B_i$, where $B_i$ is a $d \times d$ matrix. So we have the prior of $h$ as the form: $p(h) \sim \mathcal{CN}(0, D)$. Note that due to the overlapping of these blocks, $D$ is no longer a block diagonal. It has the structure that each $\gamma_i B_i$ lies along the principal diagonal of $D$ and overlaps other $\gamma_i B_i$. Starting with this model we can develop algorithms for estimating the hyper-parameters $\lambda_i$, $\lambda_i B_i$ ($\forall i$). However, because the covariance matrix $D$ is no longer a block diagonal matrix we cannot directly use the BSBL algorithm.

To facilitate the use of the BSBL framework, we expand the covariance matrix $\tilde{D}$ as follows:

$$\tilde{D} = \text{Bdiag}\{\gamma_1 B_1, \cdots, \gamma_i B_i, \cdots, \gamma_c B_c\}, \quad (8)$$

where Bdiag$\{\cdot\}$ denotes a block diagonal matrix operator with principal diagonal blocks given by $\{\gamma_1 B_1, \cdots, \gamma_c B_c\}$. Note that now $\gamma_i B_i$ does not overlap other $\gamma_j B_j$ when $i \neq j$. The expanded covariance matrix $\tilde{D}$ implies the decomposition of $h$

$$h = \sum_{i=1}^{c} E_i f_i, \quad (9)$$

where $E\{f_i\} = 0$, $E\{f_i f_j^T\} = \delta_{ij} \gamma_i B_i$ for $i = j$; $\delta_{ij} = 0$ for others), and $f \triangleq [f_1^T, f_2^T, \cdots, f_c^T]^T \sim \mathcal{CN}(0, \tilde{D})$. $E_i$ is a $N \times d$ zero matrix except that the part from its $i$-th row to $(i + d - 1)$-th row is replaced by the identity matrix $I$. Then the original system model (1) can be rewritten as

$$\bar{y} = \sum_{i=1}^{c} XE_i f_i + \bar{z} = Af + \bar{z}, \quad (10)$$

where $A \triangleq [A_1, A_2, \cdots, A_c]$ and $A_i \triangleq XE_i$. Hence, the BCS-CE method without considering cluster-partition information is summarized in Tab. 2, where $\mu_f^i$ is the corresponding $i$-th block in $\mu_f$, and $D_f^i$ is the corresponding $i$-th main diagonal block in $\tilde{D}_f$.

TABLE II. PROPOSED BCS-CE METHOD WITHOUT USING CLUSTER-PARTITION INFORMATION.

| |
|---|
| Input: $A$, $\bar{y}$, $F$ |
| Output: $\tilde{f}$ |
| Initialize: $\mu_f \leftarrow 0$, $D_f \leftarrow 0$, <br> For $i = 1, 2, \ldots, c$ <br> $\quad B \leftarrow \frac{1}{c} \sum_{i=1}^{c} \frac{D_f^i + \mu_f^i(\mu_f^i)^T}{\gamma_i}$ <br> $\quad \gamma_i \leftarrow \frac{1}{d} \text{Tr}\left[B^{-1}(D_f^i + \mu_f^i(\mu_f^i)^T)\right]$ <br> $\quad \mu_f \leftarrow \tilde{D}A^T(\lambda I + A\tilde{D}A^T)^{-1}\bar{y}$ <br> $\quad D_f \leftarrow \tilde{D} - \tilde{D}A^T(\lambda I + A\tilde{D}A^T)^{-1}A\tilde{D}$ <br> $\quad \lambda \leftarrow \frac{\|\bar{y} - A\mu_f\|_2^2 + \lambda\left[cd - \text{Tr}(D_f^i \tilde{D}^{-1})\right]}{N}$ <br> End <br> $\tilde{f} \leftarrow \tilde{D}A^T(\lambda I + A\tilde{D}A^T)^{-1}\bar{y}$ |

## V. COMPUTER SIMULATIONS

In this section, the proposed BCS-CE estimator using 100 independent Monte-Carlo runs for averaging. The length of channel vector $h$ is set as $L = 100$. Values of dominant channel taps follow multivariable Gaussian distribution and their cluster-positions are randomly allocated within the length of $h$ which is subjected to $E\{\|h\|_2^2\} = 1$, where $E\{\}$ denotes expectation operator. Please notice that all of channel taps are included in existing clusters and there is no independent channel tap in this simulation. Of course, even if there has some independent channel taps, the proposed method can still work well but corresponding performance gain is not obviously. In addition, the received signal-to-noise ratio (SNR) is defined as $10\log(E_b/\sigma_n^2)$, where $E_b = 1$.

### A. MSE versus SNR

The estimation performance is evaluated by average mean

square error (MSE) standard which is defined as

$$\text{Average MSE}\{\tilde{h}\} = E\left\{\|h - \tilde{h}\|_2^2\right\}, \quad (11)$$

where $h$ and $\tilde{h}$ are the actual channel vector and its channel estimator, respectively.

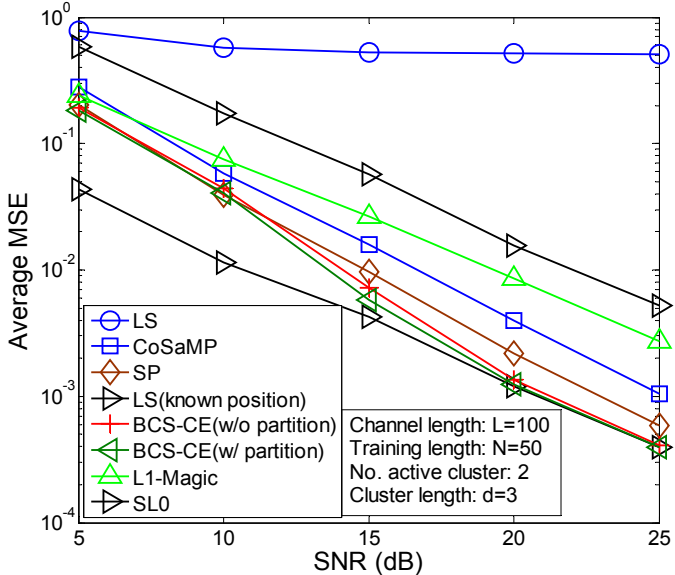
Fig. 2. Performance comparisons v.s. SNR ($N$=50 and $d$=3).

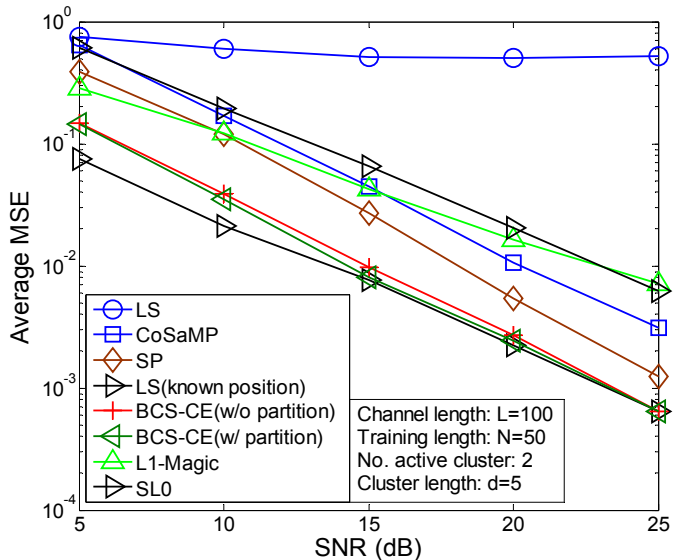
Fig. 3. Performance comparisons v.s. SNR ($N$=50 and d=5).

Considering different clusters size ($d$) as well as different training length ($N$), MSE performance is evaluated in Figs. 2-4. It is worth mentioning that the cluster size ($d$) is deterministic but unknown for channel estimation. We compare the average MSE performance of the proposed BCS-CE estimators with traditional sparse channel estimators. The lower bound is given by least square (LS) method which utilized the channel position information. In these figures, it is easy to find that proposed methods obtained lower MSE performance than traditional methods. Moreover, the proposed method achieved much better performance gain than traditional methods for bigger cluster size channels, such as in Fig. 2 ($d$=3) and Fig. 3 ($d$=5). In other words, our proposed methods can exploit more channel information for bigger cluster size. In addition, comparing with BCS-CE with utilizing cluster partition information, BCS-CE without considering the partition information, but the estimation of the latter is very close to the former one. Hence, the latter method is more practical to utilize in real communication systems.

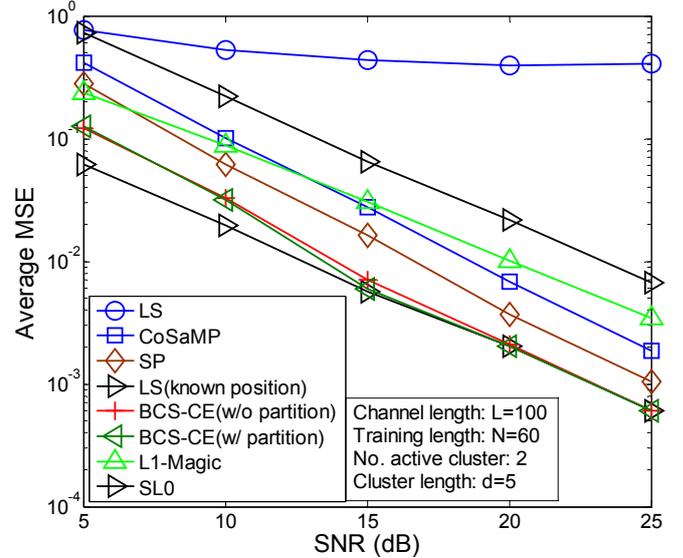
Fig. 4. Performance comparisons v.s. SNR ($N$=60 and d=$5$).

## B. Complexity evauation

To compare the computational complexity of the proposed method with other methods, CPU time is adopted for evaluation standard as shown in Fig. 5 and Fig. 6 with respect to different number of nonzero taps. It is worth mentioning that although the CPU time is not an exact measure of complexity, it can give us a rough estimation of computational complexity. Our simulations are performed in MATLAB 2012 environment using a 2.90GHz Intel i7 processor with 24GB of memory and under Microsoft Windows 7 64 bit operating system. The two figures show that the complexity of the proposed methods is close to CoSaMP which is also the low-complexity algorithm. The complexity of BCE-CE with utilizing the partition information is lower than BCE-CE without utilizing the partition information due to the latter method spent some computation for finding the active clusters.

As for the benchmarks, several conventional low-complexity sparse algorithms are adopted. From the two figures, we can find that the proposed methods are slight higher complexity than state-of-arts sparse channel estimations while achieve much better estimation performance than these methods. As the fast development of computing ability, it may not problem for achieving better performance while at the cost of amount of computation complexity.

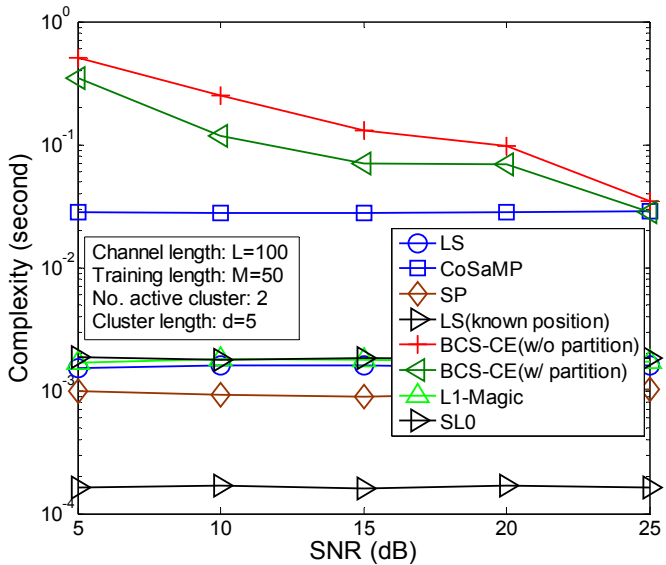

Fig. 5. Complexity comparisons via CPU time (second).

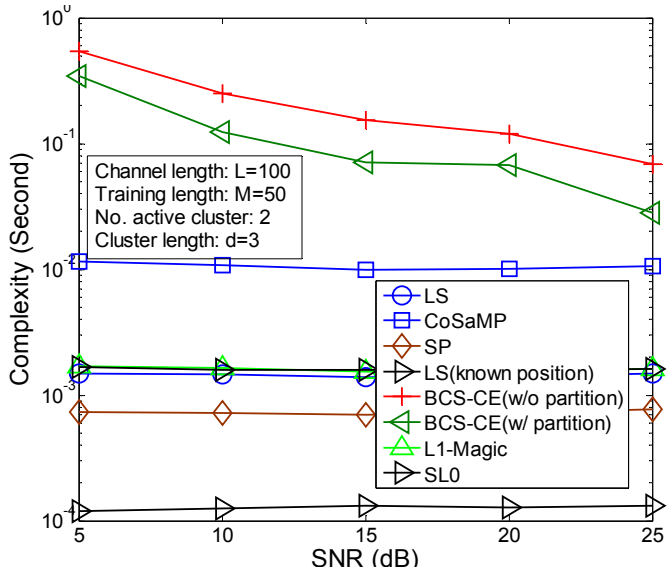

Fig. 6. Complexity comparisons via CPU time (second).

## VI. CONCLUSION AND FUTURE WORK

Traditional sparse channel estimation methods cannot exploit the inherent cluster-structure information without partition in OFDM channels. In this paper, we proposed BCS-CE methods for the estimating cluster sparse channels with or without considering a priori knowledge on cluster partition in OFDM systems. Comparing with traditional methods, the proposed methods achieved much better estimation performance due to the fact that they took advantage of potential intra-cluster correlation. Computer simulation results confirmed the effectiveness of the proposed methods but at the cost of amount of computational complexity. In future study, we are going to reduce the complexity of the proposed methods. Also, we will extend these methods to estimating channels in massive multiple-input multiple-output (massive MIMO) systems.